\documentclass[letterpaper,english,aps,manuscript]{revtex4}
\usepackage[T1]{fontenc}
\usepackage[latin1]{inputenc}
\usepackage{amsmath}
\usepackage{amssymb}

\makeatletter
\usepackage{amssymb}

\usepackage{graphicx}
\usepackage{geometry}

\usepackage{babel}

\usepackage{babel}
\makeatother
\begin{document}

\title{Hyperbolic heat equation in Kaluza's magnetohydrodynamics}

\author{A. Sandoval-Villalbazo}

\author{A. L. García-Perciante}

\address{Departamento de Física y Matemáticas, Universidad Iberoamericana,
Prolongación Paseo de la Reforma 880, México D. F. 01210, México.}

\author{L. S. García-Colín}

\address{Departamento de Física, Universidad Autónoma Metropolitana-Iztapalapa,
Av. Purísima y Michoacán S/N, México D. F. 09340, México. Also at
El Colegio Nacional, Luis González Obregón 23, Centro Histórico, México
D. F. 06020, México.}

\date{\today{}}

\begin{abstract}
This paper shows that a hyperbolic equation for heat conduction can
be obtained directly using the tenets of linear irreversible thermodynamics
in the context of the five dimensional space-time metric originally
proposed by T. Kaluza back in 1922. The associated speed of propagation
is slightly lower than the speed of light by a factor inversely proportional
to the specific charge of the fluid element. Moreover, consistency
with the second law of thermodynamics is achieved. Possible implications
in the context of physics of clusters of galaxies of this result are
briefly discussed.
\end{abstract}
\maketitle

\section{introduction\label{sec:introduction}}

It is today a well established fact that knowledge of magnetohydrodynamics
(MHD) is essential in our understanding of a vast number situations
which occur in astrophysics \cite{kul}. Less known is the relationship
which exists between the basic equations of MHD with non-equilibrium
thermodynamics, specially in the case in which one wishes to include
dissipative effects into the scheme. Four years ago, two of us \cite{pop}
wrote a paper showing clearly that when this is the case, the classical
non-relativistic structure of linear irreversible thermodynamics (LIT)
restricts the constitutive equations relating the electrical current
to an electric field to the well-known Ohm's equation when the second
law is satisfied. Non-ohmic effects can thus be included by attempting
an extension of LIT to include such cases, but a theory with these
characteristics which is also at grips with the second law of thermodynamics
does not yet exists \cite{a2}. The other alternative is to approach
the problem by incorporating the effects of the electromagnetic field
using a relativistic multidimensional theory, a step that also brings
the theory to the framework of general relativity. Many multidimensional
theories have been formulated since Einstein's idea of unifying nature's
fundamental forces in a single theory was put forward. One of the
pioneers in this topic was the mathematician T. Kaluza \cite{KK}.
Although he was not the first one to propose an extended metric, his
idea of adding extra elements to the $4\times4$ general relativity
space-time can be considered as the origin of a variety of multidimensional
theories that has been proposed in the last five decades. In this
work we formulate the thermo-hydrodynamics for an ionized fluid in
the context of Kaluza's original five-dimensional theory. This framework,
which does not account for weak and strong interactions, has sufficient
information to include gravitation and electromagnetism.

In Ref.\,\cite{pop} it was shown how the curvature arising from
the extra dimension accounts for the electromagnetic effects in the
hydrodynamic equations. This was achieved by using Meixner's theory
of non-equilibrium thermodynamics \cite{degroot,doclibro,lgcsart}
and the $5\times5$ metric proposed by Kaluza. In that work the entropy
production was calculated and it was shown how, using this formalism,
the constitutive equation for the electric current can be other than
Ohm's law, which as we know is not always valid. However, the causality
issue in the heat equation was not addressed.

In this work we complete the calculation in Ref.\,\cite{pop} and
add a new fundamental element in Kaluza's magnetohydrodynamics. A
parameter $\xi$ is introduced in the metric which ultimately adds
non-negligible corrections in the transport equations. The causality
issue in the transport equations is also addressed by calculating
the heat equation, which \emph{is shown to be of the hyperbolic type,}
and thus does not allow heat waves propagating with a speed larger
than the speed of light.

Section \ref{sec:Kformalism} is a brief summary of Kaluza's formalism.
In Sect.\,\ref{sec:dis} we establish the entropy production and
the heat equation showing how causality is not violated if the five
dimensional formalism is considered. This is accomplished by using
the standard projector $h_{\nu}^{\mu}$ \cite{eck} which eliminates
the somewhat delicate concept of dissipation in time \cite{jnet}.
A summary and a brief discussion of the results here obtained is included
in Sect.\,\ref{sec:sum}. In Appendix A some relevant details of
Kaluza's theory are shown including the relevance of the parameter
$\xi$.

\section{kaluza's formalism\label{sec:Kformalism}}

The theory formulated by T. Kaluza \cite{KK}, unifies general relativity
and electromagnetism by extending the 4-dimensional manifold into
a 5-dimensional space-time. In this metric, the fifth column contains
the components of the electromagnetic potentials:\begin{equation}
g_{\mu5}=\left[\begin{array}{c}
A_{1}\xi\\
A_{2}\xi\\
A_{3}\xi\\
\frac{\phi}{c}\xi\\
1\end{array}\right],\label{1}\end{equation}
 where $A_{\mu}$ are the components of the vector potential and $\phi$
is the electrostatic scalar potential. The constant $\xi$ is a parameter
whose value is set equal to\begin{equation}
\xi=\sqrt{\frac{16\pi G\epsilon_{0}}{c^{2}}},\label{1.5}\end{equation}
 by matching Einstein's field equation to a Poisson equation for $\phi$,
in the non-relativistic limit, as is shown in Appendix A. For simplicity
we will neglect the magnetic field and assume that the 4-manifold
is given by a quasi-Minkowski metric. That is, if the gravitational
potential is $\psi$, the metric tensor can be written as\begin{equation}
g_{\mu\nu}=\left[\begin{array}{ccccc}
1 & 0 & 0 & 0 & 0\\
0 & 1 & 0 & 0 & 0\\
0 & 0 & 1 & 0 & 0\\
0 & 0 & 0 & -1-\frac{2\psi}{c^{2}} & -\frac{\phi}{c}\xi\\
0 & 0 & 0 & -\frac{\phi}{c}\xi & 1\end{array}\right],\label{2}\end{equation}
 and\begin{equation}
g^{\mu\nu}=\left[\begin{array}{ccccc}
1 & 0 & 0 & 0 & 0\\
0 & 1 & 0 & 0 & 0\\
0 & 0 & 1 & 0 & 0\\
0 & 0 & 0 & -1+\frac{2\psi}{c^{2}} & -\frac{\phi}{c}\xi\\
0 & 0 & 0 & -\frac{\phi}{c}\xi & 1\end{array}\right],\label{3}\end{equation}
 neglecting terms of order $1/c^{3}$ and lower. In this framework,
the Christoffel symbols are proportional to the components of the
forces and\begin{equation}
\frac{dx^{5}}{dt}=\frac{q}{m\xi},\label{4}\end{equation}
 so that the equation of motion for a charged particle in a gravitational
and electric field corresponds to a geodesic in this space-time as
shown in Appendix A. It is also required that quantities have no variation
in the fifth dimension by setting $\partial/\partial x^{5}=0$. This
is called the {}``cylindrical condition'' \cite{KK} and follows
the idea that the extra dimension is closed on itself as is now assumed
in some multi-dimensional theories.

\section{heat conduction in kaluza's magnetohydrodynamics\label{sec:dis}}

As shown in Ref.\,\cite{pop}, the MHD equations can be obtained
within the five-dimensional theory. The effects of the electrical
force arise in a natural way from the curvature of space-time. No
external forces are introduced since the components of the electromagnetic
field are included in the equations through the Christoffel symbols
in the covariant derivatives (see Appendix A).

The conservation of particles is unaffected by the fifth dimension.
That is, if $n$ is the local particle density

\begin{equation}
\dot{n}+n\theta=0,\label{13}\end{equation}
 where $\theta=u_{;\mu}^{\mu}$ is the divergence of the velocity
for which the last term, $u_{,5}^{5}$, vanishes due to the cylindrical
condition. For simplicity, we will assume that the calculations are
performed in the comoving frame. Thus, $\theta=0$ and the particle
conservation law is given by $\dot{n}=0$.

The stress tensor within this framework is given by \cite{pop}\begin{equation}
T_{\nu}^{\mu}=\rho u^{\mu}u_{\nu}+ph_{\nu}^{\mu}+\Xi_{\nu}^{\mu},\label{14}\end{equation}
 where $\rho$ is the mass-energy density, $p$ the hydrostatic pressure
and $\Xi_{\nu}^{\mu}$ is the viscous stress tensor. The spatial projector
$h_{\nu}^{\mu}$ is defined in the standard way so that \begin{equation}
u^{\nu}h_{\nu}^{\mu}=0,\label{14.5}\end{equation}
 and, as will be seen,\begin{equation}
u_{\mu}\Xi_{\nu}^{\mu}=0.\label{14.6}\end{equation}
 This is achieved by having \begin{equation}
h_{\nu}^{\mu}=\delta_{\nu}^{\mu}+\frac{1}{\alpha^{2}}u_{\nu}u^{\mu},\label{15}\end{equation}
 where $\alpha^{2}=c^{2}-u_{5}u^{5}$. Some authors include a heat
term in the definition of the stress tensor following Eckart's formalism
\cite{eck}. In Meixner's theory, heat is viewed as a non-mechanical
form of energy and is included in a separate equation. A discussion
on the difference between both formalisms and a possible way of discerning
the correct way to treat heat in general relativity via a light scattering
experiment can be found in Refs.\,\cite{oviedo,jnet}. In Meixner's
thermodynamics, one considers the total energy flux: \begin{equation}
J_{[T]}^{\mu}=u^{\nu}T_{\nu}^{\mu}+n\, e_{int}u^{\mu}+J_{[Q]}^{\mu},\label{16}\end{equation}
 where $u^{\nu}T_{\nu}^{\mu}$ and $e_{int}u^{\mu}$ are the mechanical
and internal energy fluxes respectively and $J_{[Q]}^{\mu}$ is the
heat flux. Total energy conservation requires $J_{[T];\mu}^{\mu}=0$
which, using the form of the stress tensor given by Eq.\,(\ref{14}),
yields \begin{equation}
\left(u^{\nu}T_{\nu}^{\mu}\right)_{;\mu}=u_{;\mu}^{\nu}\Xi_{\nu}^{\mu}.\label{17}\end{equation}
 To obtain Eq.\,(\ref{17}) we used that, according to Einstein's
field equations, $T_{\nu;\mu}^{\mu}=0$ and that $u^{\mu}u_{\mu}=\alpha^{2}$
is a constant. Thus, in the comoving frame $\left(u^{\mu}u_{\mu}\right)_{;\nu}=0$.
Hence, the balance equation for the internal energy is \begin{equation}
n\,\dot{e}_{int}=-u_{;\mu}^{\nu}\Xi_{\nu}^{\mu}-J_{[Q];\mu}^{\mu}.\label{18}\end{equation}
 The electric contribution to the internal energy is, as mentioned
above, contained in the covariant derivatives of Eq.\,(\ref{18}).
It is pertinent to point out that Eq.\,(\ref{18}) would differ from
its Eckart´s type counterpart by a term proportional to the hydrodynamical
acceleration. In both cases, in the comoving frame, both formalisms
would lead to Eq.\,(\ref{18}).

\subsection{Entropy production}

In Meixner's theory, the entropy production is obtained by using the
local equilibrium assumption \cite{degroot,doclibro} namely\begin{equation}
s=s\left(n,\, e_{int}\right),\label{19}\end{equation}
 for which, since $\dot{n}=0$, one can write \begin{equation}
n\dot{s}=\frac{n}{T}\dot{e}_{int},\label{19.5}\end{equation}
 or, introducing Eq.\,(\ref{18}), \begin{equation}
n\dot{s}+\left(\frac{J_{[Q]}^{\mu}}{T}\right)_{;\mu}=-\frac{1}{T}u_{;\mu}^{\nu}\Xi_{\nu}^{\mu}-\frac{J_{[Q]}^{\mu}}{T^{2}}T_{,\mu}.\label{20}\end{equation}
 One can identify the right side of Eq.\,(\ref{20}) with Clausius'
uncompensated heat or the entropy production which, according to the
second law of thermodynamics is required to be non-negative:\begin{equation}
-\frac{J_{[Q]}^{\mu}}{T^{2}}T_{,\mu}-u_{;\mu}^{\nu}\Xi_{\nu}^{\mu}\geq0.\label{21}\end{equation}
 This requirement suggests constitutive equations as expressions of
each source term in Eq.\,(\ref{20}) as a product of a thermodynamic
force and its corresponding flux. Following this reasoning we write
a Fourier law for the heat flux:\begin{equation}
J_{[Q]}^{\mu}=-\kappa h_{\nu}^{\mu}T^{,\nu}.\label{22}\end{equation}
 In Eckart's theory a second term, proportional to the acceleration,
is included in Eq.\,(\ref{22}) which is not present here. This is
not only because $u^{i}=0$ in the comoving frame but primarily since,
as mentioned above, the heat flux is treated separately in Meixner's
formalism \cite{oviedo}. Anyway, \emph{both formalisms yield the
same expressions for the entropy production, heat flux and transport
equations in the comoving frame}.

In order to guarantee that the second term in Eq.\,(\ref{21}) is
positive we write\begin{equation}
\Xi_{\nu}^{\mu}=-\eta\sigma_{\nu}^{\mu},\label{23}\end{equation}
 where $\eta$ is a {}``viscosity'' coefficient. Equation (\ref{23})
relates the viscous stress tensor with the symmetrical part of the
(traceless) velocity gradient $\sigma_{\nu}^{\mu}$. We want to remark
that Eq.\,(\ref{23}) in the comoving frame is equivalent to the
standard constitutive equation $\Xi_{\nu}^{\mu}=-\eta h_{\alpha}^{\mu}h_{\nu}^{\beta}\sigma_{\beta}^{\alpha}$.
Bulk viscosity has been neglected. A calculation for $\sigma_{\nu}^{\mu}$
in the comoving frame leads to (see Appendix A): \begin{equation}
\sigma_{4}^{i}=u_{;4}^{i}=u_{;i}^{4}=\frac{1}{c}\left(\psi_{,i}+\frac{q}{2m}\phi_{,i}\right)\qquad\textrm{and}\qquad\sigma_{5}^{i}=u_{;5}^{i}=-u_{;i}^{5}=\frac{1}{2}\xi\phi_{,i},\label{23.5}\end{equation}
 for $i=1,\,2,\,3$ and the rest of the coefficients vanish. Since
in Einstein's equation, $G_{\mu\nu}$is symmetrical, the viscous stress
tensor does not have antisymmetrical components and thus we can assume
that the coupling constant which would correspond to a rotational
viscosity vanishes. Then, \begin{equation}
\Xi_{j}^{i}=0\qquad\textrm{for }i,\, j=1,\,2,\,3,\,5,\label{24}\end{equation}
 and \begin{equation}
\Xi_{i}^{4}=-\frac{\eta}{c}\left(\psi_{,i}+\frac{q}{2m}\phi_{,i}\right).\label{25}\end{equation}
 When these constitutive equations are substituted in Eq.\,(\ref{20}),
the entropy balance equation reads\begin{equation}
n\dot{s}+\left(\frac{J_{[Q]}^{\mu}}{T}\right)_{;\mu}=\kappa h_{\nu}^{\mu}\frac{T^{,\nu}T_{,\mu}}{T^{2}}+\eta\frac{1}{c^{2}}\left(\psi_{,i}+\frac{q}{2m}\phi_{,i}\right)\left(\psi^{,i}+\frac{q}{2m}\phi^{,i}\right).\label{26}\end{equation}
 The second term in the right side of Eq.\,(\ref{26}) is due to
the fields present in the problem and is obtained here as a consequence
of the curvature of space-time.

\subsection{Heat equation}

The heat equation is obtained by using the local equilibrium assumption
for the internal energy \begin{equation}
n\dot{e}_{int}=nC_{v}\frac{\partial T}{\partial t},\label{27}\end{equation}
 together with Eq.\,(\ref{18}) and the corresponding constitutive
equations given by Eqs.\,(\ref{23}) and (\ref{25}). The first three
terms in the divergence of the heat flux give the standard result\begin{equation}
\left(J_{[Q]}^{i}\right)_{;i}=-\kappa\nabla^{2}T\quad\textrm{for }i=1,\,2,\,3.\label{28}\end{equation}
 Note that, in the standard theory, the heat flux only contains the
three terms in Eq.\,(\ref{28}) which, when introduced in Eq.\,(\ref{27}),
yields a parabolic heat equation. However, in the five-dimensional
theory, the fourth component of the heat flux is nonzero retaining
its consistency with the orthogonality condition in Eq.\,(\ref{14.5}).
That is, since\begin{equation}
h_{i}^{4}=0\quad\textrm{for }i=1,\,2,\,3\label{28.5}\end{equation}
\begin{equation}
h_{4}^{4}=1-\frac{c^{2}}{\alpha^{2}},\label{28.75}\end{equation}
and by means of the cylindrical condition $T^{,5}=0$, the fourth
component of the heat flux is\begin{equation}
J_{[Q]}^{4}=-\kappa\left(1-\frac{1}{1-\delta^{2}}\right)T^{,4},\label{29}\end{equation}
 where $\delta^{2}=\frac{1}{16\pi\epsilon_{0}G}\frac{q^{2}}{m^{2}}$.
Taking the covariant derivative in Eq.\,(\ref{29}) leads to\begin{equation}
\left(J_{[Q]}^{4}\right)_{;4}=-\kappa\left(1-\frac{1}{1-\delta^{2}}\right)\left(T^{,4}\right)_{;4}=\frac{\kappa}{c^{2}}\left(1-\frac{1}{1-\delta^{2}}\right)\left(\frac{\partial^{2}T}{\partial t^{2}}-\psi_{,i}T^{,i}\right).\label{29.5}\end{equation}
 The nonlinear term $\nabla\psi\cdot\nabla T$ couples the gravitational
field to the temperature gradient. It is worth mentioning that there
is also a coupling of order $1/c^{4}$ between the electrostatic field
and the temperature gradient which consistently has been neglected.
An analysis of the consequences of these couplings is an interesting
topic which will be addressed in the future. With these extra terms,
Eq.\,(\ref{27}) can be written as \begin{equation}
D_{T}\frac{\partial T}{\partial t}=\nabla^{2}T-\frac{1}{c^{2}}\left(1-\frac{1}{1-\delta^{2}}\right)\left(\frac{\partial^{2}T}{\partial t^{2}}-\psi_{,i}T^{,i}\right)+\frac{\eta}{\kappa}\frac{1}{c^{2}}\left(\psi_{,i}+\frac{q}{2m}\phi_{,i}\right)\left(\psi^{,i}+\frac{q}{2m}\phi^{,i}\right),\label{30}\end{equation}
 where $D_{T}=nC_{v}/\kappa$ is the thermal diffusivity. Equation
(\ref{30}) is a hyperbolic, and thus causal, heat equation. Since
the coefficient $\eta$ plays the role of a viscosity, the factor
that multiplies the dissipative source on the third term on the right
side can be interpreted as a Prandtl number that quantifies the ratio
between viscous and thermal dissipations. Diffusion dominates for
small values of this parameter and, assuming the fluid is neutral,
the standard heat equation is recovered. Otherwise, since $\epsilon_{0}G$
is a very small quantity, even a very small net specific charge will
yield $\delta^{2}>1$. That is, because of the large value of $1/\xi$
the corrections due to the electric charge of matter, and thus to
the fifth dimension, are always significant unless $q/m$ is identically
zero.

The speed of propagation for the wave-like solution of Eq.\,(\ref{30})
is lower than the speed of light. If $\tilde{c}$ is the modified
speed and since $\delta^{2}\gg1$ we obtain\begin{equation}
\tilde{c}^{2}=c^{2}\left(1-\frac{1}{\delta^{2}}\right).\label{31}\end{equation}
 As an example, consider an element of fluid purely constituted by
electrons. In that case, the specific charge is $e/m_{e}$ and thus\begin{equation}
\tilde{c}^{2}=c^{2}\left(1-10^{-42}\right).\label{31}\end{equation}
 Having $\delta^{2}\neq0$ is the key element for having causal heat
conduction, although the correction imposed on the speed of propagation
is small. Therefore, this will not make significant changes in the
measurements where one would obtain $\tilde{c}\simeq c$.

\section{summary and discussion\label{sec:sum}}

The main objective of this paper is to show how using the unifying
ideas set forth by Kaluza over eighty years ago and the tenets of
LIT one can derive transport equations in MHD which incorporate not
only electrical dissipation but other dissipative effects. This is
clearly exhibited in Eq.\,(\ref{30}), the main result of this paper
in which the heat equation is modified in two ways. Firstly, it is
a hyperbolic equation which therefore does not violate causality and
is at grips with the second law of thermodynamics. Secondly, it includes
coupling effects between the external and temperature fields together
with viscosity. Thus, a relativistic framework allows for the inclusion
of dissipation consistently with a non negative entropy production.
Clearly, in the limit when $c\rightarrow\infty$ one recovers the
standard equation for heat conduction.

Finally, it is worth mentioning that the usefulness of these results
is not merely conceptual. There is an increasing interest in dealing
with the inclusion of dissipative effects in the physics of intracluster
media \cite{b1} and/or clusters of galaxies \cite{b2} when temperatures
are close to $10^{8}$ K. The existence of significant magnetic fields
in the intracluster medium is a fact that encourages further applications
of this formalism of relativistic MHD.

This work has been supported by CONACyT project 41081-F and
FICSAC, M\'{e}xico (PFSA 2005-2006).

\appendix

\section{kaluza's theory and the parameter $\xi$\label{xi}}

The Christoffel symbols are defined as\begin{equation}
\Gamma_{\beta\gamma}^{\alpha}=\frac{g^{\alpha\eta}}{2}\left(\frac{\partial g_{\beta\eta}}{\partial x^{\gamma}}+\frac{\partial g_{\eta\gamma}}{\partial x^{\beta}}-\frac{\partial g_{\beta\gamma}}{\partial x^{\eta}}\right).\label{A1}\end{equation}
 Using the metric tensor defined in Eqs.\,(\ref{1}) and (\ref{2}),
with $A_{i}=0$ one can easily obtain that\begin{equation}
\Gamma_{45}^{i}=\frac{\xi}{2c}\phi_{,i}\qquad\Gamma_{44}^{i}=\frac{1}{c^{2}}\psi_{,i}.\label{A2}\end{equation}
 The {}``shortest path'' in this five dimensional space corresponds
to a geodesic which is given by\begin{equation}
\frac{dx^{i}}{dt^{2}}+c^{2}\Gamma_{\alpha\beta}^{i}u^{\alpha}u^{\beta}=0.\label{A3}\end{equation}
 It is easy to show that using the elements in Eq.\,(\ref{A2}) and
$u^{5}=q/m\xi$, Eq.\,(\ref{A3}) corresponds precisely to the equation
of motion for a charged particle that moves in a region where the
gravitational and electrostatic potentials are $\psi$ and $\phi$
respectively. Also notice that, since the covariant derivative is
defined by \begin{equation}
V_{;\mu}^{\nu}=\frac{\partial V^{\nu}}{\partial x^{\mu}}+\Gamma_{\eta\mu}^{\nu}V^{\eta},\label{A4}\end{equation}
 for any tensor $V^{\nu}$, the components of the gradient of the
potentials will appear in a natural way in the equations. For example,
when computing a given component of the velocity gradient in the comoving
frame \begin{equation}
u_{;4}^{i}=\Gamma_{\eta4}^{i}u^{\eta}=\Gamma_{44}^{i}u^{4}+\Gamma_{45}^{i}u^{5},\label{A5}\end{equation}
 which, when the relations given in Eq.\,(\ref{A2}) are introduced,
yields Eq.\,(\ref{23.5}).

The electromagnetism equations in this framework can be obtained from
the field equations. That is \begin{equation}
R_{44}-\frac{1}{2}Rg_{44}=\kappa T_{44},\label{5}\end{equation}
 yields the Poisson equation for the gravitational potential with
$\kappa=\frac{8\pi G}{c^{4}}$ while \begin{equation}
R_{45}-\frac{1}{2}Rg_{45}=\kappa T_{45},\label{6}\end{equation}
 corresponds a Poisson equation for the electrostatic potential. As
usual, $R_{\mu\nu}$ represents the Ricci tensor. The coefficient
$\xi$ is fixed by comparing this equation with\begin{equation}
\nabla^{2}\phi=-\frac{\rho}{\epsilon_{0}}.\label{8}\end{equation}
 This part of the calculation is subtle and deserves a closer look.
In Kaluza's original work, the curvature scalar $R$ in Eq.\,(\ref{6})
is neglected. However, calculating $R_{45}$ and using the relation\begin{equation}
R=-\kappa T_{\mu}^{\mu},\label{9}\end{equation}
 one obtains, for the field equation,\begin{equation}
\frac{\xi}{2c}\nabla^{2}\phi+\frac{1}{2}\kappa\rho\left(c^{2}-\frac{q^{2}}{m^{2}\xi^{2}}\right)\frac{\phi}{c}\xi=-\frac{\kappa}{\xi}c\rho_{e},\label{10}\end{equation}
 where $\rho_{e}\equiv\rho q/m$ is the electric charge density. In
this Helmholtz-type equation, the second term corresponds to the scalar
curvature term in Einstein's field equation. Rearranging terms, one
can compare this linear term with the source on the right side as
follows. By assuming that the second term can be neglected one can
obtain\begin{equation}
\xi=\sqrt{\frac{16\pi G\epsilon_{0}}{c^{2}}}\sim6\times10^{-19}\,\frac{kg}{c}\cdot\frac{s}{m}.\label{11}\end{equation}
 Introducing this estimate for $\xi$ in Eq.\,(\ref{10}) yields\begin{equation}
\nabla^{2}\phi\sim\left(\frac{q}{m}\frac{\phi}{c}-2c\right)\frac{\kappa c}{\xi^{2}}\rho_{e},\label{12}\end{equation}
 where we neglected the first term in the parenthesis in Eq.\,(\ref{10})
since $c^{2}\ll\frac{q^{2}}{m^{2}\xi^{2}}\sim\frac{q^{2}}{m^{2}}\times3\times10^{36}\left(\frac{kg}{c}\cdot\frac{s}{m}\right)^{2}$.
Under the assumption that $\phi/c$ is small, one can also neglect
the linear term in $\phi$ and the standard Poisson equation for the
electrostatic potential is obtained.


\begin{thebibliography}{10}
\bibitem[1]{kul}{}``Plasma Physics for Astrophysicists'', Russel M. Kulsrud, Princeton
University Press, Princeton N.J. (2005).
\bibitem[2]{pop}A. Sandoval-Villalbazo and L. S. García-Colín, Phys. Plasmas \textbf{7},
4823 (2000).
\bibitem[3]{a2}L.S. García-Colín, Mol. Phys. \textbf{86}, 697 (1995).
\bibitem[4]{KK}{}``Modern Kaluza-Klein Theories'', edited by T. Appelquist, A.
Chodos, and P. Freund pp. 61 (Addison-Wesley, Reading, 1987).
\bibitem[5]{degroot}S. R. de Groot and P. Mazur, \textit{Nonequilibrium Thermodynamics}
(Dover, New York, 1984).
\bibitem[6]{doclibro}L. S. García-Colín and P. Goldstein, \textit{La Física de los Procesos
Irreversibles} (El Colegio Nacional, México 2003). (in Spanish)
\bibitem[7]{lgcsart}L. S. García-Colín, Rev. Mex. Fís. \textbf{34}, 344 (1988) (in english).
\bibitem[8]{eck}C. Eckart, Phys. Rev. 58, 919 (1940).
\bibitem[9]{jnet}L. S. García-Colín and A. Sandoval-Villalbazo; J. Non Equilibr. Thermodyn.
\textbf{31}, 1 (2006). [gr-qc/0503047]
\bibitem[10]{oviedo}A. Sandoval-Villalbazo and L. S. García-Colín, to appear in the proceedings
of the XXVIII Spanish Relativity Meeting, AIP Conference
Proceedings (2006). [gr-qc/0511055]
\bibitem[11]{b1}A.C. Fabian, C.S. Reynolds, G.B. Taylor and R.J.H. Dunn; MNRAS (2006)
to be published. [astro-ph/0501222].
\bibitem[12]{b2}M. Brüggen and M. Ruszkowsky (2005) [astro-ph/0512148].
\end{thebibliography}
\end{document}